\documentstyle[epsf]{elsart}
\newcommand {\be}{\begin{equation}}
\newcommand {\ee}{\end{equation}}
\newcommand {\bea}{\begin{eqnarray}}
\newcommand {\eea}{\end{eqnarray}}

\begin{document}
 \begin{frontmatter}
\title{Varying Alpha and the Electroweak Model}
%\title {Implication of varying alpha to the electroweak model}

\author{Dagny Kimberly and Jo\~ao Magueijo}
\address{Theoretical Physics, The Blackett Laboratory,
 Imperial College, Prince Consort Road, London, SW7 2BZ, U.K.
}

\maketitle

\begin{abstract}
Inspired by recent claims for a varying fine structure constant,
alpha, we investigate the effect of ``promoting coupling constants
to variables'' upon various parameters of the standard model. We
first consider a toy model: Proca's theory of the massive photon.
We then explore the electroweak theory with one and two dilaton
fields. We find that a varying alpha unavoidably implies varying
$W$ and $Z$  masses. This follows from gauge invariance, and is to
be contrasted with Proca' theory.
For the two dilaton theory the Weinberg angle is also variable,
but Fermi's constant and the tree level fermion masses
remain constant unless the Higgs' potential becomes dynamical. We
outline some cosmological implications.
\end{abstract}
\end{frontmatter}
%\pacs{PACS Numbers: *** }

\section{Introduction}

There is currently much interest in cosmological theories where
the conventional ``constants'' of Nature may actually vary in
space and time. The most observationally sensitive of
these``constants'' is the electromagnetic fine structure constant,
$\alpha$.  The new observational many-multiplet technique of Webb
et al, has provided the first evidence that the fine structure
constant may change throughout cosmological
time\cite{murphy,webb,webb2}. The trend of these results is that
the value of $\alpha $ was lower in the past, with $\Delta \alpha
/\alpha =-0.72\pm 0.18\times 10^{-5}$ for redshifts $z\approx
0.5-3.5.$ Other
investigations have claimed preferred non-zero values of $\Delta \alpha $ $%
<0 $ to best fit the cosmic microwave background (CMB) and Big
Bang Nucleosynthesis (BBN) data at $z\approx 10^3$ and $z\approx
10^{10}$ respectively~\cite{avelino,bat}.

A varying fine structure constant (defined to be $\alpha =e^2/4
\pi \hbar c$) may be interpreted as a varying electric charge in a
theory where $\hbar$ and $c$ are held fixed. A simple varying $e$
theory may be set up by {\it prescribing} that $e$ become a
dynamic field, the so-called minimal coupling
prescription~\cite{bek2}. This electromagnetic varying $e$ theory,
reviewed here in Section~\ref{review}, has been thoroughly
explored~\cite{bek2,bsm,olive,zal,moffatal,wep}, and a formal
rearrangement shows that it is a particular type of dilaton
theory. It is a theory in which the dilaton (a massless and gauge
neutral scalar that interacts with matter at strengths comparable
to that of gravity) couples to the electromagnetic ``$ F^2$'' term
in the Lagrangian, but not to the other gauge
fields~\cite{bsm,joaohaav}.

Given that we already know that electromagnetic and weak intereactions
are unified, a natural question is how  this electromagnetic theory
extends to the standard model of particle physics, which rests on
various ``constants'' in addition to $e$.  The sheer multitude of
``arbitrary'' parameters within the standard model has been a
source of displeasure among theorists. Thus, in considering the
electroweak extension of the electric model of~\cite{bek2},
we are led to wonder whether some
of these parameters become variables, and which are independent. 
Similar issues, in
the context of the QCD, grand-unification, and the quantum vacuum
energy, have been considered
in~\cite{land,lang,gut,dine,dine1}.

In this paper we extend the work of~\cite{bek2,bsm}
and promote the couplings in the electroweak theory to dynamical
fields. In preparation for this, in Section~\ref{nonab} we consider
general non-abelian gauge groups with a varying
coupling, and in Section~\ref{proca} Proca's theory. In the
latter, a ``gauge'' boson acquires mass by explicitly  breaking
gauge invariance. It is possible to
simultaneously have a varying $e$ and a constant boson mass in
this case. We then propose 
a version of the electroweak theory in which the $SU(2)$ gauge
charge, $g'(x^\mu)$,
in addition to the $U(1)$ gauge charge, $g(x^\mu)$, become
dynamical according to a prescription similar to the one used
in~\cite{bek2,bsm}. Again, a simplifying formal rearrangement
converts the theory into a dilaton theory, this time with two
dilaton fields that couple to the $SU(2)$ and $U(1)$ gauge fields.
A single dilaton variation is also considered.

We find that the variable couplings inevitably lead to a theory in
which the $W$ and $Z$ masses vary. This is to be contrasted with
Proca' theory and is directly related to gauge invariance. In the two
dilaton case the Weinberg angle becomes a variable too. However,
Fermi's constant and the tree level fermion masses remain
constant, unless we also promote to variables the parameters in the
Higgs' potential. We outline some
astrophysical and cosmological consequences.

\section{Varying electromagnetic alpha and dilaton theories}
\label{review} In the varying $\alpha$ theories proposed in
\cite{bek2,bsm} one 
attributes variations in $\alpha $ to changes in $e$, or the
permittivity of free space. This is done by letting $e$ take on
the value of a real scalar field which varies in space and time
$e_0\rightarrow e(x^\mu)=e_0\epsilon (x^\mu ),$ where
$\epsilon(x^\mu)$ is a dimensionless scalar field and $e_0$ is a
constant denoting the present value of $e(x^\mu)$. One then
proceeds to set up a theory based on the principles of local gauge
invariance, causality, and the scale invariance of the
$\epsilon$-field.

Since covariant derivatives take the form $D_\mu
\phi=(\partial_\mu +ieA_\mu)\phi$, for gauge transformations of
the form $\delta\phi=-i\chi \phi$ one must impose $\epsilon
A_\mu \rightarrow \epsilon A_\mu +\chi _{,\mu }$. The
gauge-invariant electromagnetic field tensor is then
\be \label{fmunu}
F_{\mu
\nu }={ (\epsilon A_\nu )_{,\mu }-(\epsilon A_\mu )_{,\nu }\over
\epsilon} ,
\ee
which reduces to the usual form for constant
$\epsilon $. The electromagnetic Lagrangian density is still \be{\cal L}
_{em}=-{F^{\mu \nu }F_{\mu \nu }\over 4}, \ee and the dynamics of
the $\epsilon $ field are controlled by the kinetic term \be{\cal
L}_\epsilon =-\frac 12 \omega {\epsilon _{,\mu }\epsilon ^{,\mu
}\over {\epsilon ^2}}, \ee
where the coupling constant $\omega$ is introduced into the
lagrangian density for dimensional reasons and is proportional to
the inverse square of the characteristic length scale of the
theory, $\omega\sim\ell^{-2}$, such that $\ell\geq
L_p\approx10^{-33}cm$ holds \cite {bek2}. This length scale
corresponds to an energy scale ${\hbar c \over \ell}$, with an
upper bound set by experiment.  Note that the metric signature
used is $(-,+,+,+)$.

A simpler formulation of this theory~\cite{bsm} can be constructed
by defining an auxiliary gauge potential $a_\mu \equiv \epsilon
A_\mu ,$ and field tensor $f_{\mu \nu }\equiv\epsilon F_{\mu \nu
}=\partial _\mu a_\nu -\partial _\nu a_\mu .$ The covariant
derivative then assumes the familiar form, $D_\mu =\partial _\mu
+ie_0a_\mu $, and the dependence of the Lagrangian on $\epsilon $
occurs only in the kinetic term for $\epsilon $ and in the
$F^2=f^2/\epsilon ^2$ term, not in the rest of the matter
lagrangian ${\cal L}_m$ (where it could only have appeared via the
covariant derivative $D_\mu$). To simplify further, we can redefine
the variable, $\epsilon \rightarrow \psi \equiv ln\epsilon .$ The
total action then becomes \be
S=\int d^4x\left({\cal L}_{mat}-{\frac \omega 2}\partial _\mu 
\psi \partial ^\mu \psi -\frac {1}{4}e^{-2\psi}f_{\mu \nu }f^{\mu \nu }
\right) ,\ee 
 where
the matter Lagrangian ${\cal L}_{mat}$ does not contain $\psi$.
This is a dilaton theory coupling to the electromagnetic ``$f^2$''
part of the Lagrangian only.  Note the scale invariance of
the action and $\psi$ (that is, their invariance under the transformation 
$\epsilon \rightarrow k\epsilon $ for any constant $k$).

Given this mathematical trick, one may wonder which of the two
sets of variables are the physics ones? The question is obviously
irrelevant regarding $A_\mu$ and $a_\mu$, because both are
unphysical due to gauge invariance. On the other hand both
$F_{\mu\nu}$ and $f_{\mu\nu}$ are ``physical'' and may be used as
convenient (the problem is similar to the use of field $\mathbf E$
or displacement $\mathbf D$ in dielectric electrostatics.) Note
that the homogeneous Maxwell equations,
$\epsilon^{\mu\nu\alpha\beta}\partial_\nu f_{\alpha\beta}=0$, are
not valid for $F_{\alpha\beta}$.

\section{Varying couplings for non-abelian gauge groups}
\label{nonab} The tools derived for electromagentism carry over
trivially to non-abelian groups (see~\cite{land} for an example
based on QCD). We take as an example $O(3)$. Let ${\mathbf \Phi}$
be a 3-vector, with covariant derivative \be D_\mu{\mathbf \Phi}
=\partial_\mu {\mathbf \Phi} +g' {\mathbf W}_\mu \wedge {\mathbf
\Phi} .\ee Here the gauge boson ${\mathbf W}_\mu$ is a 3-vector.
Under a gauge transformation corresponding to a  rotation defined
by vector ${\mathbf \Xi}$, we have: \bea \delta{\mathbf
\Phi}&=&-{\mathbf \Xi} \wedge {\mathbf
\Phi},\\
g'\delta {\mathbf W}_\mu &=&\partial_\mu {\mathbf \Xi} -{\mathbf
\Xi} \wedge g' {\mathbf W}_\mu ,\eea
 Written in this form,
these equations are preserved even if $g'$ becomes variable,
$g'\rightarrow g'(x^\mu)=\eta (x^\mu)g_0 $. The field tensor is
now \be \label{fmununa}
{\mathbf W}_{\mu\nu}={1\over g'}[\partial_\mu (g'{\mathbf
W}_\nu) -
\partial_\nu (g'{\mathbf W}_\mu)] + g' {\mathbf W}_\mu\wedge
{\mathbf W}_\nu ,\ee so that it is covariant, \be \delta {\mathbf
W}_{\mu\nu}= - {\mathbf \Xi} \wedge {\mathbf W}_{\mu\nu}, \ee and
a possible Lagrangian is \be {\cal L}_{\mathbf W}=-{1\over
4}{\mathbf W}_{\mu\nu}\cdot {\mathbf W}^{\mu\nu}.\ee

As before we can define an auxiliary gauge boson $g'{\mathbf
W}_\mu\equiv g'_0 {\mathbf w}_\mu$ and an auxiliary field
$g'{\mathbf W}_{\mu\nu}\equiv g'_0 {\mathbf w}_{\mu\nu}$, or
equivalently $\eta (x^\mu){\mathbf W}_\mu \equiv {\mathbf w}_\mu$
and $\eta(x^\mu){\mathbf W}_{\mu\nu}\equiv{\mathbf w}_{\mu\nu}$.
With these definitions, the field $\eta(x^\mu)$ does not appear in
the gauge derivative \be D_\mu{\mathbf \Phi} =\partial_\mu
{\mathbf \Phi} +g'_0 {\mathbf w}_\mu \wedge {\mathbf \Phi}, \ee
and thus not in the matter Lagrangian. The gauge Lagrangian
becomes \be {\cal L}_{\mathbf w}=-{1\over
4}e^{-2\chi}{\mathbf w}_{\mu\nu}\cdot {\mathbf w}^{\mu\nu} \ee
with $\chi\equiv\log (g'(x^\mu)/g'_0)$, or $g'(x^\mu)=g_0'e^\chi$.
As we can see, with a couple of trivial modifications, the tools
previously developed for the $U(1)$ gauge group may be adapted to
any non-abelian gauge theory.

\section{The Proca theory and explicit breaking of gauge invariance}
\label{proca}

Another interesting extension of the varying $e$ electromagnetic
model of~\cite{bek2,bsm} is Proca's theory of the massive photon.
It will prove useful as a contrast to the electroweak
results, where gauge bosons acquire a mass via a quite different
mechanism.

The Proca lagrangian, with a dynamic electromagnetic coupling
given by $e(x^\mu)=e_oe^\psi$, may be written as: \be {\cal
L}_P={\cal L}_m-{\omega\over 2}\psi_{,\mu}\psi^{,\mu}-{1\over
4}F_{\mu\nu}F^{\mu\nu}-{1\over 2}m^2A_\mu A^\mu,\ee for a photon
with mass $m$, assumed to be a constant parameter. The covariant
derivative appearing in ${\cal L}_m$ is, say,  $D_\mu
\phi=(\partial_\mu +ieA_\mu)\phi$ for transformations of the form
$\delta\phi=-i\chi \phi$ and $\epsilon A_\mu \rightarrow \epsilon
A_\mu +\chi _{,\mu }$. Even though the mass term breaks gauge
invariance it is still possible to define a gauge-invariant
electromagnetic field tensor according to (\ref{fmunu}).

As before we can define $f_{\mu \nu}=\partial_\mu
a_\nu-\partial_\nu a_\mu$, with $a_\mu \equiv \epsilon A_\mu ,$
leading to Lagragian:
\be {\cal L}_P=  {\cal L}_m-{\omega\over
2}\psi_{,\mu}\psi^{,\mu}-{1\over
4}e^{-2\psi}f_{\mu\nu}f^{\mu\nu}\nonumber-{1\over 2}e^{-2\psi}
m^2a_\mu a^\mu,\ee
where the matter lagrangian now does not contain $\psi$ (since the
covariant derivative becomes $D_\mu \phi=(\partial_\mu
+ie_0a_\mu)\phi$). In terms of these variables it is easy to find
that the dynamical equation for $\psi$ is
\be\label{psiproca} \Box\psi=-{1\over\omega}({1\over
2}F_{\mu\nu}F^{\mu\nu}+m^2A_\mu A^\mu). \ee

With loss of gauge invariance, the question of which of $A_\mu$ or
$a_\mu$ is the physical field acquires relevance. From the $a_\mu$
formulation it seems that the photon mass has to be variable in
this theory; the opposite conclusion is reached from the $A_\mu$
formulation. We can quickly see, however, that a variable mass is
not the correct physical interpretation. Varying the lagrangian
with respect to, say,  $A^\mu$ leads to Maxwell's equations
\be \epsilon \partial_\mu {F^{\mu\nu}\over \epsilon}-m^2
A^\nu=J^\nu \ee
The equation of current conservation is now $\partial_\mu
(J^\mu/\epsilon)=0$, so that one must have $\partial_\mu
(A^\mu/\epsilon)=0$. 
The wave equation in free space is therefore
\be \label{aproca}(\Box - m^2) A^\nu - ...
%F^{\mu\nu}\partial_\mu\psi
=0 \ee
where the ellipse refers to terms in $\partial_\mu\psi$. 
One may check by direct substitution into (\ref{psiproca}) and
(\ref{aproca}) 
that there are plane wave solutions for all
reasonable amplitudes and wavelengths (note that $m^2 \ll
\omega$). Their dispersion relations (for either $A^\mu$ or
$a^\mu$ waves) are $E^2-{\mathbf k}^2=m^2$, that is, the photon
has a constant mass $m$ regardless of the field variable used.
This is also the mass that appears in the propagator for this
theory. More generally the physical mass should be identified 
from the Lagrangian written in terms of variables
such that the gradient terms have no prefactor.

The conclusion is that a varying charge does not imply a varying
boson mass if the latter is obtained by explicitly breaking gauge
invariance. Of course we can, if we wish, also have a varying
photon mass, by promoting $m$ to a dynamical field; but this is
not necessary. The situation will be different for the vector
boson masses in the electroweak extension of the Bekenstein model,
where gauge invariance is preserved.

\section{The electroweak model}

We are now ready to consider the electroweak sector of the
standard model~\cite{gl,we,sa}. Its fundamental degrees of freedom
are massless spin 1/2 chiral particles $\Psi_i$, and the gauge
symmetry group is $SU(2)_L\otimes U(1)$, where $SU(2)$ is weak
isospin (acting on left handed fermions only) and $U(1)$ is the
weak hypercharge. The coupling constants are $g_0$ and $g_0'$ for
the $U(1)$ and $SU(2)$ interactions, respectively.

As before, we promote the  gauge couplings to
fields, writing  $g'(x^\mu)=\eta'(x^\mu)g'_0$ and
$g(x^\mu)=\eta(x^\mu)g_0$. We may then define  fields $\psi$ and
$\chi$ via:
 \bea
g'(x^\mu)&=&g'_0e^\psi ,\\
g(x^\mu)&=&g_0e^\chi .
\eea
Again, we may avoid the presence of $\psi$ and $\chi$ in
covariant derivatives by
defining auxiliary
gauge boson fields, \bea
g'(x^\mu){\mathbf W}_\mu&=&g'_0{\mathbf w}_\mu ,\\
g(x^\mu) Y_\mu&=&g_0y_\mu .\eea
Then, considering for example the Higgs field (a
complex doublet $\Phi$), the derivative:
\be D_\mu\Phi= \left (\partial_\mu -{i\over 2}g' {\bf
t}\cdot{\mathbf W}_\mu -{i\over 2}g Y_\mu  \right) \Phi ,\ee
(where ${\bf t}$ are the SU(2) generators) becomes
\be D_\mu\Phi= \left
(\partial_\mu -{i\over 2}g'_0 {\bf t}\cdot{\mathbf w}_\mu -{i\over
2}g_0 y_\mu  \right) \Phi .\ee
We may also define field
tensors:
\bea \label{gaugeL}{\mathbf w}_{\mu\nu}&=&\partial_\mu {\mathbf
w}_\nu -
\partial_\nu
{\mathbf w}_\mu - g'_0{\mathbf w}_\mu \wedge{\mathbf w}_\nu,\\
y_{\mu\nu}&=&\partial_\mu y_\nu - \partial_\nu y_\mu, \eea
or similar expression for ${\mathbf W}_{\mu\nu}$ and $Y_{\mu\nu}$, written
in terms of ${\mathbf W}_\mu$ and $Y_\mu$ (see (\ref{fmunu}) and 
(\ref{fmununa})).

The core electroweak Lagrangian may now be written as ${\cal L}={\cal
L}_{{\mathbf w}y}+{\cal L}_\Phi+{\cal L}_{\psi\chi}$. The gauge
field Lagrangian is:
\bea {\cal L}_{{\mathbf w}y }&=&-{1\over 4}{\mathbf
W}_{\mu\nu}\cdot{\mathbf W}^{\mu\nu} -{1\over
4}Y_{\mu\nu}Y^{\mu\nu}\nonumber\\
&=&-{1\over 4} e^{-2\psi}{\mathbf w}_{\mu\nu}\cdot {\mathbf
w}^{\mu\nu}-{1\over 4} e^{-2\chi}y_{\mu\nu}y^{\mu\nu},\eea
and (as in Proca's theory) using variables ${\mathbf W}^\mu$ and
$Y^\mu$ (in terms of which the gauge Lagrangian has no prefactor)
facilitates identifying the physical masses that appear in the
dispersion relations and in the propagators. The Higgs Lagrangian
is
\be {\cal L}_\Phi=(D_\mu \Phi)^\dagger(D_\mu \Phi)-V(\Phi),\ee
with potential
\be V(\Phi)={m^2\over
2}\Phi^\dag\Phi+{\lambda\over4}(\Phi^\dag\Phi)^2 \ee
As in Proca's theory with varying electric charge, the potential
parameters $m$ and $\lambda$ may be assumed to be constant.
Finally the fields $\chi$ and $\psi$ acquire dynamics via
\be {\cal L}_{\psi\chi} =-{\omega'\over
2}\psi_{,\mu}\psi^{,\mu}-{\omega\over 2}\chi_{,\mu}\chi^{,\mu}
.\ee
We could also consider a simpler, one-dilaton variation of this
theory by identifying $\chi$ and $\psi$, and keeping just one of
the terms in ${\cal L}_{\psi\chi}$.

In order to induce spontaneous symmetry breaking of the $SU(2)$
gauge group we should choose $m^2<0$. Then the potential has a
minimum at $|\Phi|^2=v^2_0\equiv\ {-m^2/\lambda} \neq0$, so the
vacuum state may be at $(\Phi)_0=\left(
\begin{array}{c}0\\v_0\end{array} \right).$
Given that the symmetry is local, a perturbative expansion around
the vacuum can always be written as
\be  \Phi(x^\mu)=\left(
\begin{array}{c}0\\v_0+{\sigma(x^\mu)\over\sqrt{2}}\end{array} \right).
\ee
One may now expand the Lagrangian. The crucial term for
identifying the boson masses is the Higgs' gradient term:
 \bea \label{grad}(D_\mu\phi)^\dag
(D^\mu\phi)&=&{1\over 2}(\partial_\mu\sigma)^2  +{{v_0}^2\over
4}(g')^2[({W_\mu}^1)^2+({W_\mu}^2)^2]\nonumber\\
&+&{{v_0}^2\over 4}(g'{W_\mu}^3-g Y_\mu)^2.\eea
From this expression we define a massless gauge field, $A_\mu$,
and its orthogonal field, $Z_\mu$, with respect to the fields
$W^3_\mu$ and $Y_\mu$
\bea Z_\mu&{\equiv}&{g'W_{\mu}^3-gY_\mu\over
\sqrt{{g'}^2+g^2}}= \cos\theta_W W_{\mu}^3-\sin\theta_W Y_\mu,\nonumber \\
A_\mu &{\equiv}& {gW_{\mu}^3+g'Y_\mu\over \sqrt{{g'}^2+g^2}} =
\sin\theta_W W_{\mu}^3+\cos\theta_W Y_\mu\label{rot}
 ,\eea
where $\theta_W$ is the weak mixing angle, or Weinberg angle given
by
\be \tan\theta_W={g\over g'}={g_0\over g'_0}e^{\chi-\psi} \ee
In the two-dilaton theory this is a variable. We could have
defined a similar (but constant) rotation in terms of the fields
$w^3_\mu$ and $y_\mu$, and this would still diagonalize the mass
matrix for a Lagrangian (\ref{grad}) rewritten in terms of these
variables. However such a rotation would induce photon-$Z$ couplings
in the kinetic terms (\ref{gaugeL}) and therefore would not be the
correct Weinberg rotation. Thus the Weinberg angle must be
variable in the two-dilaton theory. Obviously in the single
dilaton theory (where $\chi$ and $\psi$ are identified) $\theta_W$
remains constant.

Once this rotation is performed we find the following tree-level
masses for the gauge bosons:
\bea m_{W}&=&{v_0\over {\sqrt 2}}g'\propto  e^\psi  \\
{m_Z}&=&{v_0 \over {\sqrt 2}}\sqrt{g'^2+g^2}={v_0 \over
{\sqrt 2}}\sqrt{g_0'^2e^{2\psi}+g_0^2e^{2\chi}}\\
m_A&=&0\eea
where we used the charged $W^\pm_\mu=(W^1_\mu\pm i W^2_\mu)/{\sqrt
2} $. We shall not discuss in this paper radiative corrections to
these formulae.

The variability of these masses is to be contrasted with Proca'
theory. There, mass and charge are essentially independent and so
it is possible to have a constant photon mass and a varying
electric charge. In the standard model, on the other hand, gauge
invariance precludes an explicit mass term. Gauge bosons acquire a
mass because they couple to the ``charged'' Higgs field via the
covariant derivative and the Higgs field undergoes spontaneous
symmetry breaking. Thus, the gauge bosons have mass only because
the Higgs field has charge, and a varying charge necessarily
implies a concomitant varying gauge boson mass.

\section{Lepton charges and masses}
We now consider the leptonic sector of the theory. For the sake of
brevity we will consider  only the electron and the electron
neutrino, but our considerations can be easily extended to
include muon and tau leptons, as well as quarks. The left handed
fermions are placed in weak isospin doublets $L=\left(
\begin{array}{c}\nu_e\\e_L\end{array}\right)$ whereas the right
handed fermions are singlets $e_R$. Bearing in mind that the
covariant derivatives are:
\bea  D_\mu L&=&{\left( \partial_\mu -{\imath\over
2}g'{\mathbf\tau}\cdot{\mathbf W}_\mu +{\imath\over 2}gY_\mu\right)} L\\
D_\mu R&=&\partial_\mu R+\imath gY_\mu R,\eea
we arrive at the free fermion lagrangian
\be
 {\cal
L}_f=\imath\bar{R}\gamma^\mu(\partial_\mu+\imath g Y_\mu)R
+\imath\bar{L}\gamma^\mu(\partial_\mu+\imath g Y_\mu-{\imath\over
2}g'{\mathbf\tau}\cdot{\mathbf W}_\mu)L \ee
After rotation (\ref{rot}) this becomes:
\bea {\cal L}_f &=&\imath \bar{e}\gamma^\mu\partial_\mu
e+\imath\bar{\nu}\gamma^\mu\partial_\mu\nu-g'\sin\theta_W\bar{e}\gamma^\mu
eA_\mu\\
&+&{g'\over \cos\theta_W}(\sin^2\theta_W\bar{e}_R\gamma^\mu e_R
-{1\over 2}\cos(2\theta_W)\bar{e}_L\gamma^\mu e_L
+{1\over 2}\bar{\nu}\gamma^\mu\nu)Z_\mu\\
&+&{g'\over 2}[(\bar{\nu}\gamma^\mu e_LW^{+\dag}_\mu)+h.c.],\eea where
$h.c.$ denotes hermitian conjugate. This expression allows us to
identify the electromagnetic and weak currents. We find that the
field $A^\mu$ is indeed the electromagnetic field, and that the
electric charge is given by
\be e=g\cos\theta_W=g'\sin\theta_W={gg'\over \sqrt{g^2+g'^2}} \ee
The fine structure ``constant'' $\alpha$ is therefore fixed by a
nontrivial combination of the fields $\psi$ and $\chi$, should there
be two dilatons. In the single dilaton case this reduces to
$e=e_0e^\chi$.

One may also identify the weak currents to find the expression for
the Fermi constant. One finds
\be G_F={ \sqrt 2 \over 4}{g'^2\over M_W^2}={1\over 2\sqrt 2
v_0^2} \ee
Interestingly, this does not vary. Fermi's constant is determined
by the Higgs' potential only, and so, for as long as its
parameters are held fixed, varying couplings in the standard model
do not lead to a varying Fermi constant.

Finally we consider the Higgs-fermion interaction Lagragian,
through which fermions acquire their masses once the Higgs
acquires a vacuum expectation value. This may be written as
 \bea
{\cal
L}_{\Phi_{int}}&=&-G_e(\bar{L}\Phi R+\bar{R}\Phi^\dag L)\\
&=&-G_e(v_0+{\sigma(x^\mu)\over{\sqrt 2}})(\bar{e}_Le_R
+\bar{e}_Re_L),\eea
where we have used the vacuum expectation value for $\Phi$ chosen
earlier, and where $G_e$ is the Higgs-Lepton coupling strength for
the electron. The electron mass is therefore given by $m_e=v_0
G_e$. Again, if the Higgs' potential parameters are kept fixed and
the parameters $G_i$ are not promoted to dynamical variables, the
tree level fermion masses remain constant even if the couplings
$g$ and $g'$ are promoted to fields.

\section{Equations of motion and  applications}
We reserve to a future publication a complete study of the
cosmological and astrophysical implications of this theory, but
here we outline some areas of interest. 

The Einstein's equations
for this theory are:
\be G_{\mu\nu}= 8\pi G(T_{\mu\nu}^{EW'} e^{-2\psi}
+T_{\mu\nu}^{EW}e^{-2\chi} +T_{\mu\nu}^{\psi} +T_{\mu\nu}^{\chi}
+T_{\mu\nu}^{mat}),\ee
that is, one must add the stress energy tensor of fields $\chi$
and $\psi$ to the right hand side. In addition we have
 \bea \Box\psi &=&
-{1\over 2\omega '} e^{-2\psi}{\mathbf w}_{\mu\nu}\cdot {\mathbf w}^{\mu\nu},\\
\Box\chi &=& -{1\over 2\omega}e^{-2\chi}y_{\mu\nu}y^{\mu\nu}.\eea
For the single dilaton theory one identifies $\psi$ and $\chi$ and
the last two equations are replaced by
\be \Box\chi = -{1\over 2\omega}e^{-2\chi}({\mathbf
w}_{\mu\nu}\cdot {\mathbf w}^{\mu\nu}+ y_{\mu\nu}y^{\mu\nu}). \ee
We can analyze these equations for two general cases. The first is
for spatially-varying, time-independent coupling fields $\psi$ and
$\chi$, for which we can find a spherically symmetric solution to
the equations of motion (an extension of the considerations
in~\cite{bt,covvsl,stars,wep}). These can then be applied to
scenarios in which weak interactions are non-negligible, such as
around massive objects like neutron stars and black holes. The
second case is for time-varying, spatially-independent fields
$\psi$ and $\chi$, which is applicable to cosmological scenarios
(an extension of the work in~\cite{bsm}).

From this exercise we may expect that the Webb results imply
significant variations in the $W$ and $Z$ masses and in the Weinberg
angle in the very early universe, in neutron stars, or near 
black holes and their accretion disks. This has obvious implications
for the physics of neutron stars, BBN,
and the electroweak phase transition.
But perhaps the most dramatic implication may be the stability of 
solitonic solutions in the standard model. Semi-local strings are 
defects that owe their stability 
to non-topological considerations~\cite{semil}. They are present
in the electroweak theory, and their region of stability
has been studied~\cite{elect,elect1}. It appears that this
region does not include the parameter values observed in the 
``actual'' standard model. However, according
to the theory presented in this paper, these parameter values are 
not constants of Nature. It is conceivable that the region of stability
for electroweak strings may be realized
in the very early universe or near neutron stars.

\section{Conclusions}\label{conc}
In this paper we examined the implications of a varying alpha in
the light of the electroweak theory. We already know that
electromagnetism and weak interactions are unified. Hence a
varying alpha {\it implies} variability for the two coupling
``constants'' of the electroweak theory. These variations may be
controlled by one or two independent ``dilaton'' fields.

We found that with coupling variability, the gauge boson
masses must also vary. This conclusion is
hardly surprising and can be qualitatively understood. In Proca's
theory an explicit mass term is added to the ``photon'' Lagrangian,
thereby breaking gauge invariance. This mass term is independent
of the charge couplings and so it is possible to accommodate both
a varying electromagnetic coupling and a constant ``photon'' mass.

The origin of the $W^\pm$ and $Z$ masses is quite different. In
the standard model, gauge invariance is fully preserved, and gauge
bosons have mass because the Higgs field undergoes spontaneous
symmetry breaking. But more important, {\it the gauge bosons only
have mass because the Higgs field carries charge}, that is, it
couples to the gauge bosons that {\it will} acquire mass. Thus, it
is impossible to have a standard model with varying charges
without passing this variability on to the $W^\pm$ and $Z$ masses,
and ultimately also the Weinberg angle (in the two-dilaton case).

The situation is again different for the tree level lepton masses.
These are not due to charge, but to the interaction with the Higgs
field via ``Yukawa''  couplings. Unless the Higgs potential
becomes dynamical, fermion masses do not change even if the
couplings do. It would be interesting to explore a variation of
the theory proposed in this paper where the Higgs' potential
becomes dynamical and so the fermion masses can vary too. Perhaps
such a theory could explain the mystery of the fermion masses, but
this is merely a speculation.

In summary, we have explored the implications of a varying alpha for
other parameters of the standard model. If in a future experiment
we were to find that the observed variations in
alpha are not accompanied by specific variations in the $W$ and $Z$ masses we
should be very worried indeed. Such a finding would imply a
violation of gauge invariance and contradict the standard model.
If we found that the Weinberg angle did not change that would be
less apocalyptic. It would simply imply that the observed variation in
alpha is due to a single dilaton field within the framework of the
standard model. If, on the other hand, we were to find that the Fermi
constant varied, or that the fermion tree-level masses varied,
then we would know that the theory presented in this paper is
too tight a framework. We would need to ``promote to
variables'' the parameters in the Higgs potential. The Mexican
hat would have to become dynamical.

\end{document}